\documentclass{iopjournal}

\usepackage{graphicx}
\usepackage{multirow}
\usepackage{amsmath,amssymb,amsfonts}
\usepackage{amsthm}
\usepackage{mathrsfs}
\usepackage[title]{appendix}
\usepackage{xcolor}
\usepackage{textcomp}
\usepackage{manyfoot}
\usepackage{booktabs}
\usepackage{listings}
\usepackage{array}
\usepackage{stfloats}
\usepackage{url}
\usepackage{verbatim}
\usepackage{comment}
\usepackage{bm}
\usepackage{float}
\usepackage{placeins}
\usepackage{caption}
\usepackage{multicol}
\usepackage{subcaption}
\usepackage{scalerel}
\usepackage{esdiff}
\usepackage{cleveref}
\usepackage{etoolbox}
\usepackage{cases}
\usepackage{mathtools}
\usepackage[ruled,vlined,linesnumbered]{algorithm2e}
\usepackage{rotating}
\usepackage[none]{hyphenat}
\usepackage{microtype}

\begin{document}

\articletype{Paper}

\title{Two-stage Distributed Variational Quantum Eigensolver Software for  QUBO and Quadratic Programming}

\author{Milad Hasanzadeh$^1$\orcid{0000-0002-0196-1116} and Amin Kargarian$^{1,*}$\orcid{0000-0000-0000-0000}}

\affil{$^1$Department of Electrical and Computer Engineering, Louisiana State University, Baton Rouge, LA 70803, USA}

\affil{$^*$Author to whom any correspondence should be addressed.}

\email{mhasa42@lsu.edu, kargarian@lsu.edu}


\begin{abstract}
This paper proposes a two-stage distributed variational quantum eigensolver (DVQE) software for solving quadratic unconstrained binary optimization (QUBO) problems and bounded constrained quadratic programming (QP) problems. The proposed DVQE solver supports both monolithic and distributed quantum-circuit execution and evaluates QUBO objectives directly from measured bitstrings. To improve variational training, DVQE uses a two-stage procedure that combines metaheuristic warm-start initialization with sampling-based variational refinement. The software supports several metaheuristic approaches as warm-start strategies. To extend QUBO-based quantum optimization to constrained continuous problems, this paper also develops a sequential QP to QUBO framework, called QQP. QQP first scales the bounded continuous variables to a normalized box and then handles equality and inequality constraints using a Powell-Hestenes-Rockafellar (PHR) augmented-Lagrangian formulation. Under a fixed PHR active region, the constrained augmented-Lagrangian subproblem becomes an ordinary bounded quadratic problem. QQP then solves this bounded quadratic problem through repeated local one-bit QUBO reformulations, where each binary variable represents a local up/down move of one continuous variable inside a trust region. In this way, QQP converts a constrained continuous QP into a sequence of QUBO subproblems without introducing slack variables. Each local QUBO subproblem can be solved using either a classical QUBO backend or the proposed DVQE solver. Numerical experiments evaluate the proposed software on QUBO and QP test problems. The results show that the distributed DVQE framework can recover high-quality QUBO solutions, and that the QQP framework can solve bounded constrained QP instances with small optimality, feasibility, and solution gaps. The complete software and tutorial examples are publicly available on \href{https://github.com/LSU-RAISE-LAB/DVQE.git}{GitHub}.
\end{abstract}

\section{Introduction}
\label{sec:introduction}

The rapid growth in the size, dimensions, and coupling structure of modern computational problems is increasingly challenging the capabilities of classical computing architectures \cite{horowitz2019quantum,gill2022quantum}. Many real-world applications, including finance \cite{egger2020quantum}, energy and power system optimization \cite{ajagekar2019quantum,hasanzadeh2026all}, satellite image processing \cite{10339907}, and high-dimensional control and decision-making systems \cite{hasanzadeh2024dynamic,saadati2024federated}, require solving large-scale optimization, simulation, or search problems whose computational cost can grow rapidly with problem size. In many cases, these problems involve combinatorial search spaces, nonlinear interactions, integer or binary decisions, and coupled constraints, which make exact classical solutions difficult or computationally expensive. Although classical algorithms and high-performance computing platforms remain the dominant tools for practical computation, their sequential logic and binary representation can become limiting when the feasible search space grows combinatorially or when many dependent variables must be processed simultaneously. These limitations have motivated growing interest in quantum computing as a complementary computational paradigm \cite{horowitz2019quantum}. Foundational quantum algorithms, such as Grover's search algorithm, Shor's factoring algorithm, and the Harrow--Hassidim--Lloyd algorithm for linear systems, have shown that quantum computation can provide provable speedups for specific computational tasks \cite{grover1996fast,shor1994algorithms,harrow2009quantum}. On the other side, the limitations of near-term noisy quantum hardware have shifted significant attention toward hybrid quantum-classical algorithms, where parameterized quantum circuits are trained by classical optimization loops \cite{peruzzo2014variational,cerezo2021variational}.

Several quantum optimization algorithms have been developed under this general hybrid quantum-classical principle. Among them, the Quantum Approximate Optimization Algorithm (QAOA) and the Variational Quantum Eigensolver (VQE) are two of the most studied approaches \cite{farhi2014quantum,peruzzo2014variational,cerezo2021variational}. These algorithms are especially relevant to quadratic unconstrained binary optimization (QUBO) problems, where the goal is to minimize a quadratic objective function over binary decision variables. QAOA was introduced for approximate combinatorial optimization and is naturally suited to QUBO and Ising models, since the objective function can be encoded as a cost Hamiltonian and optimized through alternating cost and mixing operations \cite{farhi2014quantum}. VQE was originally proposed for estimating the ground-state energy of molecular Hamiltonians, but the same variational principle can also be applied to QUBO problems after mapping the binary objective to an Ising Hamiltonian \cite{peruzzo2014variational}. In this setting, a parameterized quantum circuit prepares a trial quantum state, measurements generate binary samples or estimate the corresponding energy, and a classical optimizer updates the circuit parameters to search for low-energy solutions. Therefore, both QAOA and VQE can be used as hybrid quantum-classical methods for QUBO-type problems. 

Although QUBO provides a convenient interface between optimization problems and near-term quantum algorithms, it is not the native mathematical form of many practical engineering problems \cite{glover2018tutorial,lucas2014ising}. A standard QUBO model is unconstrained and uses binary decision variables, whereas many real-world optimization problems include continuous variables, box limits, equality constraints, inequality constraints, and coupled physical or operational restrictions \cite{boyd2004convex,hasanzadeh2025dplib,nocedal2006numerical,hasanzadeh2025admm}. In particular, quadratic programming (QP) and linear programming (LP) models appear frequently, and their constraints cannot be directly represented in the basic QUBO form \cite{boyd2004convex}. As a result, QAOA- and VQE-based quantum optimization methods usually require the original problem to be converted into a QUBO or Ising Hamiltonian before a parameterized quantum circuit can be applied. This creates an important modeling gap: constrained continuous optimization problems require additional reformulation layers before they can be solved through a QUBO-based quantum backend \cite{glover2018tutorial,lucas2014ising}.

Even when a constrained or continuous optimization problem is successfully converted into a QUBO model, another major challenge remains: the resulting QUBO may require more qubits and circuit resources than are available on a single near-term quantum device. In VQE-based QUBO solving, each binary variable is typically represented by one qubit, and additional circuit depth is needed to create correlations among variables. As the number of binary variables increases, the required number of qubits, two-qubit gates, measurements, and classical parameter updates can grow rapidly. This makes large QUBO instances difficult to solve on a single noisy intermediate-scale quantum (NISQ) processor, where qubit counts, gate fidelities, connectivity, and coherence times remain limited \cite{preskill2018quantum,cerezo2021variational}. Distributed quantum computing has therefore emerged as a promising direction for extending quantum computation beyond the capacity of one monolithic processor by dividing a larger quantum circuit across multiple smaller quantum processing units connected through communication or circuit-distribution mechanisms \cite{caleffi2024distributed,buhrman2003distributed}. For QUBO optimization, this direction is particularly attractive because the objective value can be estimated from measured binary samples while the ansatz can be executed over a distributed multi-QPU structure.

Several recent studies have investigated distributed quantum computing from an algorithmic and architectural perspective. \cite{diadamo2021distributed} proposed a distributed framework for accelerated VQE, where the conventional sampling process is replaced by a rejection-filtering phase-estimation procedure. \cite{tan2022distributed} studied distributed quantum computation for Simon's problem and developed a distributed quantum algorithm that can be extended across multiple computing nodes. Earlier, \cite{yimsiriwattana2004distributed} presented a distributed implementation of Shor's factoring algorithm over a quantum network model. More recently, \cite{zhou2023distributed} proposed distributed exact quantum algorithms for Bernstein--Vazirani and search problems, including a distributed exact Grover-type search algorithm. These works show that distributing quantum circuits across multiple quantum processing units can reduce single-device resource requirements and support larger algorithmic structures. However, they mainly focus on specific distributed quantum algorithms, communication models, or architectural execution strategies. Although several quantum software frameworks have been published for quantum-classical programming, quantum simulation, and variational-circuit training \cite{mccaskey2018language,powers2021mistiqs,acampora2024evovaq,mccaskey2020xacc,dahlberg2019simulaqron,haner2018software}, they do not provide an end-to-end reusable software framework for distributed VQE-based QUBO solving, nor do they connect distributed quantum execution with a continuous QP to QUBO reformulation layer. This leaves an important practical gap between distributed quantum algorithm design and a testable optimization-oriented software.

Beyond modeling and hardware-scaling limitations, variational quantum optimization also faces a major training challenge. Whether the quantum circuit is executed on a single processor or distributed across multiple processors, QAOA- and VQE-type methods still depend on a classical outer-loop optimizer to tune the parameters of the ansatz. This training process is difficult because the objective landscape can be highly nonconvex, noisy, and sensitive to initialization. In particular, barren plateaus can occur when the gradient of the cost function becomes exponentially small with respect to the number of qubits, making gradient-based training ineffective for larger circuits \cite{mcclean2018barren,cerezo2021cost,larocca2025barren}. Even when gradient-free optimizers are used to avoid direct gradient estimation, there is still no guarantee that the selected ansatz and classical optimizer will find parameters close to the true ground state. Poor initialization may trap the training process in low-quality regions of the parameter space, while measurement noise and finite sampling can further distort the estimated energy values used by the optimizer \cite{cerezo2021variational}. Therefore, practical VQE-based QUBO solving requires not only a QUBO reformulation layer and a scalable circuit-execution model, but also an effective parameter-initialization and training strategy that improves the chance of finding high-quality low-energy bitstrings.

Motivated by these gaps, this paper develops an end-to-end optimization-oriented framework that connects distributed variational quantum execution, QUBO solving, and constrained continuous optimization. First, we propose a distributed variational quantum eigensolver (DVQE) for QUBO problems, where a parameterized ansatz can be executed either on a monolithic quantum processor model or over a distributed multi-QPU structure. To improve the training of the variational circuit, the proposed DVQE uses a two-stage strategy: a metaheuristic warm-start phase, based on Black Hole optimization, Grey Wolf Optimization, or Artificial Bee Colony optimization, followed by a sampling-based variational refinement phase. Second, we develop a sequential QP to QUBO reformulation framework, called QQP, that extends QUBO-based optimization to bounded constrained continuous problems. QQP first maps the original bounded variables to a normalized box, which improves numerical conditioning and gives a common scale for the local search steps. Equality and inequality constraints are then handled through a Powell-Hestenes-Rockafellar (PHR) augmented-Lagrangian formulation, where equality violations are penalized quadratically and inequality constraints are treated through a projected multiplier term. For a fixed PHR region, this piecewise quadratic augmented-Lagrangian model becomes an ordinary bounded quadratic subproblem. QQP solves this subproblem by repeatedly constructing local one-bit QUBO models, where each binary variable selects a local upward or downward move of one continuous variable within a trust region. The accepted local move updates the continuous iterate, and the augmented-Lagrangian multipliers and PHR region are then updated until feasibility and objective improvement are achieved. In this way, QQP provides the missing reformulation layer between constrained continuous optimization models and QUBO-based quantum backends. The software is available on \href{https://github.com/LSU-RAISE-LAB/DVQE.git}{GitHub}, accompanied by several tutorial examples \cite{GitHub_raiselab}.

The main contributions of this paper are summarized as follows:
\begin{itemize}
    \item A distributed variational quantum eigensolver (DVQE) is developed with an open-source software for solving QUBO problems using Qiskit Aer as the quantum-circuit simulation backend.

    \item A two-stage DVQE training strategy is proposed, combining metaheuristic warm-start with sampling-based variational refinement to improve parameter initialization and solution quality.

\item A sequential QP/LP-to-QUBO framework, called QQP, is developed to reformulate bounded constrained QP and LP problems into local one-bit QUBO subproblems that can be solved using the proposed DVQE solver.

\end{itemize}

The rest of this paper is organized as follows. Section~\ref{sec:preliminaries} reviews the required background on QP/LP models, QUBO models, metaheuristic optimization, variational quantum algorithms, and distributed quantum computing. Section~\ref{sec:proposed_framework} presents the proposed DVQE framework, the DVQE software package, the sequential QQP reformulation method, and the QQP software package. Section~\ref{sec:experiments} reports the numerical experiments, including DVQE ablation studies, comparison with Gurobi MIQP, QQP validation with a classical internal QUBO backend, and QQP experiments using DVQE as the internal QUBO solver. Finally, Section~\ref{sec:conclusion} concludes the paper.

\section{Preliminaries}
\label{sec:preliminaries}

This section provides the background needed for the proposed DVQE and QQP frameworks. It first introduces the QUBO model used by quantum optimization solvers, then reviews bounded QP and LP models, variational quantum circuits, distributed quantum execution, and the metaheuristic methods used later for warm-start initialization.

\subsection{QUBO Formulation}
\label{subsec:qubo_formulation}

Quadratic unconstrained binary optimization is a standard model for binary optimization problems and is the main problem class solved by the proposed DVQE framework. A QUBO problem is written as
\begin{equation}
\min_{z \in \{0,1\}^{N}} \quad z^{T}Qz + q^{T}z,
\label{eq:qubo}
\end{equation}
where $z$ is a binary decision vector, $Q$ contains quadratic coefficients, and $q$ contains linear coefficients. The goal is to find the binary vector with the minimum QUBO cost.

A QUBO objective can also be represented as an Ising-type Hamiltonian. Using the binary-to-spin relation
\begin{equation}
z_i = \frac{1-s_i}{2}, \qquad s_i \in \{-1,1\},
\label{eq:binary_spin}
\end{equation}
the QUBO objective can be expressed in terms of Pauli-$Z$ operators as
\begin{equation}
H_{\mathrm{QUBO}}
=
\sum_i h_i Z_i
+
\sum_{i<j} J_{ij} Z_iZ_j
+
\kappa I,
\label{eq:ising}
\end{equation}
where $h_i$ and $J_{ij}$ are determined by the QUBO coefficients, and $\kappa$ is a constant energy shift. The ground state of this Hamiltonian corresponds to the minimum-cost binary solution of the QUBO problem.

In the implementation used in this work, the Pauli representation is retained for compatibility and inspection. However, the scalable optimization routine evaluates QUBO costs directly from measured bitstrings. Therefore, a measured bitstring is treated as a candidate binary solution, and its quality is computed using the original QUBO objective in~\eqref{eq:qubo}.

\subsection{QP and LP Problems}
\label{subsec:qp_lp}

In addition to direct QUBO solving, this work considers bounded continuous optimization problems with linear constraints. A bounded quadratic program can be written as
\begin{equation}
\begin{aligned}
\min_{x \in \mathbb{R}^{n}} \quad 
& \frac{1}{2}x^{T}Hx + f^{T}x \\
\mathrm{s.t.} \quad 
& Ax \leq b, \\
& A_{\mathrm{eq}}x = b_{\mathrm{eq}}, \\
& \ell \leq x \leq u,
\end{aligned}
\label{eq:qp}
\end{equation}
where $x$ is the continuous decision vector, $H$ is the quadratic cost matrix, $f$ is the linear cost vector, $A$ and $A_{\mathrm{eq}}$ define the inequality and equality constraints, and $\ell$ and $u$ are lower and upper bounds. When $H=0$, problem~\eqref{eq:qp} reduces to a bounded linear program,
\begin{equation}
\begin{aligned}
\min_{x \in \mathbb{R}^{n}} \quad 
& f^{T}x \\
\mathrm{s.t.} \quad 
& Ax \leq b, \\
& A_{\mathrm{eq}}x = b_{\mathrm{eq}}, \\
& \ell \leq x \leq u .
\end{aligned}
\label{eq:lp}
\end{equation}

QP and LP models appear in many engineering and decision-making applications. However, they contain continuous variables and explicit equality and inequality constraints, so they are not directly compatible with QUBO-based quantum solvers. The QQP framework developed in this paper addresses this gap by transforming bounded constrained QP and LP models into a sequence of local QUBO subproblems.

\subsection{Quantum Circuits and VQE}
\label{subsec:ansatz_vqe}

A variational quantum algorithm uses a parameterized quantum circuit, or ansatz, to prepare a trial quantum state. For an $N$-qubit QUBO problem, the ansatz prepares
\begin{equation}
|\psi(\theta)\rangle
=
U(\theta)|0\rangle^{\otimes N},
\label{eq:ansatz_state}
\end{equation}
where $U(\theta)$ is a quantum circuit with tunable parameters $\theta$. A typical ansatz is built from parameterized single-qubit rotation gates and fixed entangling gates. The rotation gates control the local state of each qubit, while the entangling gates introduce correlations among qubits. Increasing the circuit depth can improve expressiveness, but it also increases the number of trainable parameters and the cost of optimization.

In the variational quantum eigensolver, the objective is to find parameters that minimize the expected energy of the Hamiltonian associated with the optimization problem,
\begin{equation}
E(\theta)
=
\langle \psi(\theta) | H_{\mathrm{QUBO}} | \psi(\theta) \rangle .
\label{eq:vqe_energy}
\end{equation}
In practice, the circuit is executed repeatedly, measurement outcomes are collected, and the energy is estimated from the measured bitstrings. A classical optimizer then updates the circuit parameters, and the process is repeated until a stopping condition is reached.

\begin{figure}[!t]
    \centering
    \includegraphics[width=0.75\linewidth]{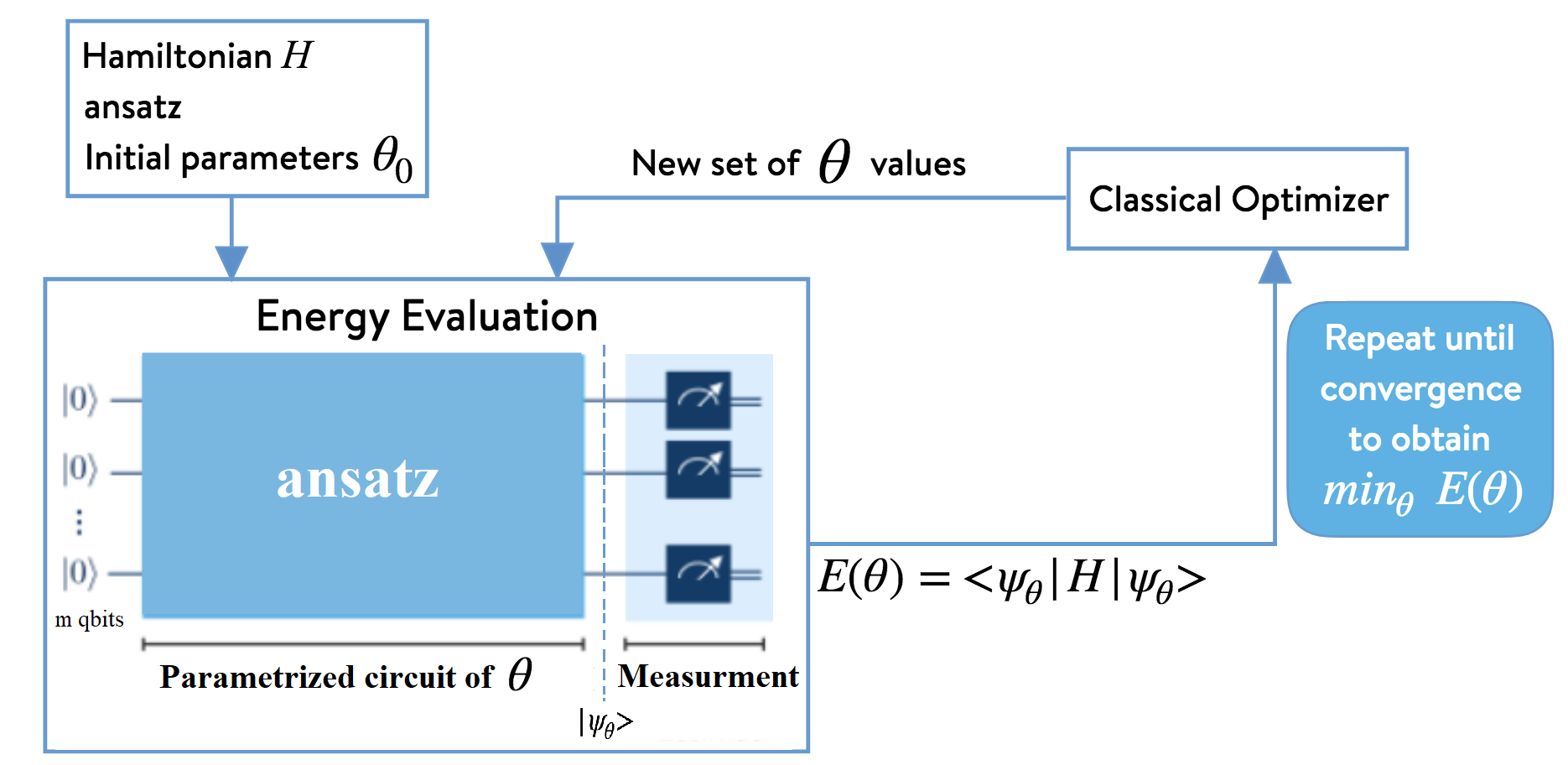}
    \caption{Closed-loop structure of the VQE workflow, showing the interaction between the parameterized quantum circuit and the classical optimizer.}
    \label{fig:vqe_loop}
\end{figure}

After training, the final circuit is sampled, and the measured bitstrings are evaluated using the original QUBO objective. The bitstring with the lowest sampled QUBO cost is selected as the candidate solution. This closed-loop structure is shown in Fig.~\ref{fig:vqe_loop}.

\subsection{Distributed Quantum Computing}
\label{subsec:distributed_quantum_computing}

Distributed quantum computing executes a quantum algorithm over multiple quantum processing units instead of a single device. This model is motivated by the limited number of reliable qubits available on near-term hardware. If a circuit requires $N$ logical qubits, these qubits can be assigned across a set of QPUs,
\begin{equation}
\mathcal{P}=\{P_1,P_2,\ldots,P_m\},
\label{eq:qpu_set}
\end{equation}
through an allocation map
\begin{equation}
\pi:\{1,2,\ldots,N\}\rightarrow\{1,2,\ldots,m\},
\label{eq:qubit_map}
\end{equation}
where $\pi(i)=j$ means that logical qubit $i$ is assigned to QPU $P_j$.

Gates acting on qubits located on the same QPU can be executed locally. The main challenge appears when a two-qubit gate acts on qubits assigned to different QPUs. Such a gate must be replaced by a distributed operation that uses communication resources, such as communication qubits, shared entanglement, measurement, classical messages, and conditional local gates. Under ideal communication assumptions, this replacement reproduces the action of the original monolithic gate, but it also introduces circuit and communication overhead.

Thus, a distributed quantum circuit requires three basic steps: assigning logical qubits to QPUs, identifying local and cross-QPU gates, and replacing cross-QPU gates with communication-compatible operations. In this work, distributed quantum computing provides the execution model for the proposed distributed VQE-based QUBO solver.

\subsection{Metaheuristic Methods}
\label{subsec:metaheuristics}

The proposed DVQE framework uses metaheuristic methods later as warm-start initialization tools for the ansatz parameters. In this setting, a candidate solution of the metaheuristic is not a binary QUBO solution. Instead, it is a vector of variational circuit parameters. Each candidate parameter vector is evaluated by running the parameterized circuit, sampling bitstrings, and estimating the corresponding QUBO energy. The best parameter vector found by the metaheuristic is then used to initialize the sampling-based VQE refinement stage.

Metaheuristic algorithms are stochastic search methods that do not require gradient information. They are useful when the objective is noisy, expensive to evaluate, nonconvex, or difficult to differentiate. These properties are relevant to variational quantum training because measured energy estimates can be noisy and the parameter landscape can contain broad flat regions. In this work, three population-based metaheuristics are used as warm-start strategies: Black Hole, Grey Wolf Optimizer, and Artificial Bee Colony.

In the Black Hole method, a population of candidate solutions is interpreted as a group of stars. The candidate with the best objective value is selected as the black hole, and the remaining candidates move toward it according to
\begin{equation}
x_i^{k+1}
=
x_i^{k}
+
r_i\left(x_{\mathrm{BH}}^{k}-x_i^{k}\right),
\label{eq:bh_update}
\end{equation}
where $x_i^k$ is the $i$th candidate at iteration $k$, $x_{\mathrm{BH}}^k$ is the current best candidate, and $r_i$ is a random number in $[0,1]$. Candidates that move too close to the black hole may be reinitialized to preserve exploration.

Grey Wolf Optimizer is inspired by the leadership hierarchy and hunting behavior of grey wolves. The three best candidates are denoted by $\alpha$, $\beta$, and $\delta$, and the rest of the population updates its position based on these leaders. In a compact form, the update can be written as
\begin{equation}
x^{k+1}
=
\frac{x_1^{k}+x_2^{k}+x_3^{k}}{3},
\label{eq:gwo_update}
\end{equation}
where $x_1^k$, $x_2^k$, and $x_3^k$ are candidate positions influenced by the current $\alpha$, $\beta$, and $\delta$ solutions. Random coefficients control the balance between exploration and exploitation, and this balance gradually shifts as the iterations proceed.

Artificial Bee Colony is based on the foraging behavior of honey bees. Candidate solutions are treated as food sources. Employed bees explore local modifications of existing food sources, onlooker bees choose promising sources based on their fitness values, and scout bees replace stagnant sources with new random candidates. This mechanism allows the method to exploit good regions while still introducing new candidates when the search becomes stagnant. All three methods are gradient-free and population-based. Therefore, they can explore the ansatz-parameter space without requiring derivative estimates.

\section{Proposed Framework}
\label{sec:proposed_framework}

This section presents the proposed framework. We first describe the distributed VQE algorithm for solving QUBO problems. The second subsection presents the DVQE software package and explains how it can be installed and used. The last two subsections present the proposed QP/LP solution framework using DVQE as a QUBO solver and the corresponding QQP software package.

\subsection{Proposed Two-stage Distributed VQE}
\label{subsec:proposed_dvqe}

The proposed DVQE method is a variational algorithm for solving QUBO problems in either monolithic or distributed quantum execution. The algorithm starts from the QUBO objective and builds a parameterized quantum circuit whose measured bitstrings represent candidate binary solutions. The circuit is then trained through two stages. The first stage performs a gradient free search over the ansatz parameter space to obtain a useful initial quantum state. The second stage refines this state through sampling based variational optimization. This design is intended to reduce the dependence on purely random initialization and to improve the probability of reaching low cost bitstrings after training. Algorithm~\ref{alg:proposed_dvqe} summarizes the procedure.

The algorithm begins with the QUBO objective $C(z)=z^TQz+q^Tz$, where each binary vector $z$ represents one candidate solution. This objective is the central quantity used by the method. The role of the quantum circuit is to generate bitstrings with high probability near low values of $C(z)$. Therefore, the algorithm does not need to form or diagonalize a full Hamiltonian matrix during training. Instead, it evaluates the quality of a parameterized circuit through the costs of the bitstrings produced by measurement.

The first line of Algorithm~\ref{alg:proposed_dvqe} constructs the parameterized ansatz $U(\theta)$. This circuit maps the all zero initial state to a trial quantum state,
\[
|\psi(\theta)\rangle = U(\theta)|0\rangle^{\otimes N},
\]
where $N$ is the number of QUBO variables. The ansatz contains parameterized single qubit rotations and entangling gates. The rotation gates provide the trainable degrees of freedom, while the entangling gates allow correlations among QUBO variables. The depth $d$ controls how many layers are used. A larger depth increases the expressive power of the ansatz, but it also increases the number of trainable parameters and the difficulty of optimization.

\begin{algorithm}[H]
\small
\caption{Distributed VQE for QUBO Problems}
\label{alg:proposed_dvqe}
\DontPrintSemicolon
\SetAlgoLined
\SetKwInOut{Input}{Input}
\SetKwInOut{Output}{Output}

\Input{QUBO objective $C(z)=z^TQz+q^Tz$, ansatz depth $d$, QPU layout, initialization rule, training budget, sampling budget}
\Output{Best sampled binary solution $z^\star$, trained circuit $U(\theta^\star)$, and final histogram $\mathcal{H}$}

Construct a parameterized ansatz $U(\theta)$ for the QUBO variables\;

Embed $U(\theta)$ into the target quantum execution model\;

Measure only the computational qubits associated with the QUBO variables\;

Create the sampled energy estimator $\widehat{E}(\theta)$ from measured QUBO costs\;

Obtain an initial parameter vector $\theta^{(0)}$ by random sampling or metaheuristic warm start\;

Set $\theta_0\leftarrow\theta^{(0)}$ and initialize the best parameter record\;

\For{each refinement iteration}{
Generate two perturbed parameter vectors around the current point\;

Estimate their sampled energies using the measured circuit\;

Compute a simultaneous perturbation descent direction\;

Update the ansatz parameters\;

Update the best parameter record if a lower sampled energy is found\;

Stop if the parameter change is smaller than the tolerance\;
}

Set $\theta^\star$ equal to the best parameter vector found during training\;

Execute $U(\theta^\star)$ with the final sampling budget and collect $\mathcal{H}$\;

Select
\[
z^\star
=
\arg\min_{z\in\mathrm{supp}(\mathcal{H})}
C(z).
\]

\Return{$z^\star$, $U(\theta^\star)$, $\mathcal{H}$}
\end{algorithm}

The second line embeds the ansatz into the target quantum execution model. In a monolithic execution model, the full ansatz is placed on one QPU. In a distributed execution model, the logical qubits of the ansatz are assigned across multiple QPUs. Local gates remain on their assigned QPUs, while gates acting on qubits located on different QPUs are represented by communication compatible operations. This step gives the algorithm its distributed form. The goal is to preserve the logical action of the variational circuit while allowing the circuit to be executed under local QPU capacity limits.

The third line adds measurements only to the computational qubits associated with the QUBO variables. The communication qubits support the execution of cross QPU operations, but they are not decision variables of the QUBO problem. Therefore, they are excluded from the measured QUBO bitstring. Each measured string used by the optimizer must correspond only to a binary vector $z\in\{0,1\}^{N}$.

The fourth line defines the sampled energy estimator $\widehat{E}(\theta)$. For a fixed parameter vector $\theta$, the circuit is executed several times and produces a histogram of measured bitstrings. The sampled energy is then computed from the QUBO costs of these bitstrings. One possible estimator is the mean sampled cost,
\[
\widehat{E}_{\mathrm{mean}}(\theta)
=
\sum_{z\in\mathrm{supp}(\mathcal{H}_{\theta})}
p_{\theta}(z)C(z),
\]
where $\mathcal{H}_{\theta}$ is the histogram produced by $U(\theta)$ and $p_{\theta}(z)$ is the empirical probability of bitstring $z$. Another useful estimator is the CVaR cost, where only the best fraction of the sampled bitstrings is used. The CVaR estimator is useful for QUBO optimization because the objective is not only to reduce the average sampled cost, but also to increase the probability of observing high quality low cost bitstrings.

The fifth line is the first stage of the proposed two stage training method. The purpose of this stage is to choose a good starting point for the variational refinement. A purely random parameter vector may place the ansatz in a poor or flat region of the objective landscape. To reduce this risk, the proposed method allows the initial parameter vector to be obtained through a gradient free metaheuristic search. In this stage, a population of candidate parameter vectors is created over the parameter domain. Each candidate is assigned to the ansatz, the circuit is measured, and its quality is evaluated using $\widehat{E}(\theta)$. The population is then updated using the selected metaheuristic rule, such as BH, GWO, or ABC. The best parameter vector found at the end of this search becomes $\theta^{(0)}$.

This warm start stage is an algorithmic part of the variational method, not a classical solution of the QUBO problem. The search is performed over the continuous ansatz parameter vector $\theta$, not over the binary decision vector $z$. Every candidate parameter vector is evaluated through circuit execution and measurement. Thus, the metaheuristic stage is used to prepare a better variational state before the main training stage begins. Since the search is gradient free, it can explore the ansatz parameter space even when local gradient information is noisy, small, or uninformative. This is why the warm start stage can empirically reduce the chance of starting the refinement inside poor flat regions of the energy landscape.

After initialization, the algorithm enters the second training stage. This stage refines the ansatz parameters using sampled energy values. At each refinement iteration, two perturbed parameter vectors are generated around the current point,
\[
\theta_k^{+}=\theta_k+c_k\Delta_k,
\qquad
\theta_k^{-}=\theta_k-c_k\Delta_k,
\]
where $\Delta_k$ is a random vector with entries in $\{-1,1\}$ and $c_k$ controls the perturbation size. Both perturbed circuits are measured, and their sampled energies are estimated. These two energy values are then used to approximate a descent direction,
\[
\widehat{g}_k
=
\frac{\widehat{E}(\theta_k^{+})-\widehat{E}(\theta_k^{-})}{2c_k}
\Delta_k .
\]
The parameters are updated as
\[
\theta_{k+1}
=
\theta_k-a_k\widehat{g}_k,
\]
where $a_k$ controls the update size.

This refinement rule is suitable for the proposed setting because it requires only two sampled energy evaluations per iteration, regardless of the number of ansatz parameters. This is important for larger QUBO instances and deeper ansatz circuits. A standard coordinate finite difference method would require a number of circuit evaluations that grows with the number of parameters. In contrast, simultaneous perturbation keeps the per iteration sampling cost fixed while still providing a useful descent direction.

During refinement, the algorithm keeps the best parameter vector observed so far. This is necessary because the sampled objective is stochastic. Due to finite shots, a later parameter vector is not always better than an earlier one, even if the general training direction is improving. Storing the best observed parameter vector makes the final trained circuit less sensitive to sampling fluctuations. The training loop stops when the parameter update becomes sufficiently small or when the training budget is exhausted.

After the refinement stage, the best recorded parameter vector is denoted by $\theta^\star$. The trained circuit $U(\theta^\star)$ is then executed with a larger final sampling budget to obtain the histogram $\mathcal{H}$. The final solution is selected by evaluating the original QUBO cost over the support of this histogram,
\[
z^\star
=
\arg\min_{z\in\mathrm{supp}(\mathcal{H})}
C(z).
\]
This final decoding rule is cost based rather than frequency based. The most frequent bitstring is not necessarily the lowest cost bitstring, especially when the sampled distribution is broad. Therefore, the algorithm returns the best QUBO solution actually observed during final sampling.

Overall, the proposed DVQE algorithm combines distributed variational circuit construction, gradient free warm start, sampled energy refinement, and cost based decoding. The distributed circuit construction allows the ansatz to be represented across multiple QPUs. The warm start stage improves the initial variational state before refinement. The simultaneous perturbation stage trains the ansatz with a fixed number of sampled energy evaluations per iteration. The final decoding step ensures that the returned bitstring is selected according to the original QUBO objective.

\subsection{DVQE Software Package}
\label{subsec:dvqe_package}

The proposed DVQE algorithm is implemented as a Python software package for solving QUBO problems using either monolithic or distributed variational quantum execution. The package provides a single high level interface while keeping the main circuit construction, distribution, training, sampling, and decoding steps modular internally. The implementation uses Qiskit for circuit construction and simulation, and Diskit for distributed circuit remapping.

The main entry point of the package is the function
\begin{verbatim}
dvqe(...)
\end{verbatim}
which executes the complete workflow. The function receives the QUBO data, execution mode, ansatz depth, initialization choice, training parameters, sampling parameters, and QPU layout for distributed execution. It returns the best sampled solution, the final trained circuit, and the final measurement histogram. A typical call has the form
\begin{verbatim}
z_best, final_circuit, hist = dvqe(
    mode="distributed",
    Q=Q,
    q_linear=q_linear,
    init_type=2,
    depth=depth,
    lr=lr,
    max_iters=max_iters,
    qpu_qubit_config=[4, 3, 2],
    rel_tol=rel_tol,
    num_shots=1024,
    final_shots=4000,
    warm_start_population=4,
    warm_start_iters=8,
    warm_start_shots=256,
    energy_mode="cvar",
    cvar_alpha=0.2
)
\end{verbatim}

The input \texttt{mode} selects the execution architecture. When \texttt{mode="monolithic"}, the ansatz is built as a standard circuit on one QPU. When \texttt{mode="distributed"}, the logical QUBO qubits are distributed across multiple QPUs according to \texttt{qpu\_qubit\_config}. The input \texttt{Q} is the quadratic QUBO matrix and \texttt{q\_linear} is the linear QUBO vector. Together, they define the objective
\[
C(z)=z^TQz+q^Tz .
\]
The input \texttt{depth} controls the number of ansatz layers. Increasing \texttt{depth} increases circuit expressiveness but also increases the number of trainable parameters.

The input \texttt{init\_type} selects the ansatz initialization strategy. The value \texttt{init\_type=1} uses random initialization. The values \texttt{init\_type=2}, \texttt{init\_type=3}, and \texttt{init\_type=4} use Black Hole, Grey Wolf Optimizer, and Artificial Bee Colony initialization, respectively. These options implement the first stage of the two stage training method. The parameters \texttt{warm\_start\_population}, \texttt{warm\_start\_iters}, and \texttt{warm\_start\_shots} control the population size, number of metaheuristic iterations, and number of shots used for evaluating candidates during warm start.

The input \texttt{lr} is the learning rate scale used in the simultaneous perturbation refinement stage. The input \texttt{max\_iters} is the maximum number of refinement iterations, while \texttt{rel\_tol} is the stopping tolerance based on the maximum parameter update. The parameters \texttt{num\_shots} and \texttt{final\_shots} control the number of circuit shots used during training and final sampling. The input \texttt{energy\_mode} selects the sampled energy estimator. When \texttt{energy\_mode="mean"}, the package uses the average QUBO cost of all measured bitstrings. When \texttt{energy\_mode="cvar"}, only the best fraction of sampled bitstrings is used, controlled by \texttt{cvar\_alpha}.

The software package is organized around a cached circuit evaluation structure. It first builds the parameterized ansatz circuit once. For distributed execution, it also builds the Diskit topology once and remaps the distributed circuit once. Measurements are added once to the computational QUBO qubits, and the measured parameterized circuit is optionally transpiled once. After this setup, every objective evaluation only binds a new parameter vector to the existing circuit template and runs the software.

The QUBO utilities provide basic cost and Hamiltonian support. The function
\begin{verbatim}
qubo_cost(z, Q, q_linear)
\end{verbatim}
computes the original QUBO objective for a binary vector. The function
\begin{verbatim}
qubo_to_pauli_hamiltonian(Q, q_linear)
\end{verbatim}
keeps the Pauli Hamiltonian representation available for compatibility and inspection. However, the scalable training loop does not require forming dense Hamiltonian matrices. Instead, sampled bitstrings are evaluated directly using the QUBO cost.

The ansatz construction functions build symbolic parameterized circuits. The function
\begin{verbatim}
create_monolithic_parameterized_ansatz(num_qubits, depth)
\end{verbatim}
constructs the monolithic ansatz. The function
\begin{verbatim}
create_distributed_parameterized_ansatz(qregs, depth)
\end{verbatim}
constructs the ansatz over distributed QPU registers. Communication registers are not directly parameterized by the ansatz because they do not represent QUBO decision variables. The function
\begin{verbatim}
bind_parameter_values(circuit, parameters, values)
\end{verbatim}
inserts numerical parameter values into a symbolic circuit.

The distributed execution support is handled through the topology and remapping functions. The function
\begin{verbatim}
prepare_topology(mode, n, qpu_qubit_config)
\end{verbatim}
creates the QPU topology for distributed execution. In monolithic mode, no distributed topology is needed. In distributed mode, the QPU configuration is checked against the number of QUBO variables, and a Diskit topology is created. The function
\begin{verbatim}
remap_with_diskit(ansatz_qc, topology)
\end{verbatim}
then remaps the circuit according to the distributed topology.

The measurement utilities ensure that only QUBO computational qubits are measured. The function
\begin{verbatim}
get_computational_qubits(circuit, mode, n)
\end{verbatim}
returns the qubits associated with the QUBO variables and excludes communication qubits in distributed mode. The function
\begin{verbatim}
add_qubo_measurements_once(circuit, mode, n)
\end{verbatim}
adds a classical register and measures only these computational qubits. The function
\begin{verbatim}
normalize_counts_to_qubo_bits(counts_raw, n)
\end{verbatim}
converts Qiskit count strings into QUBO bitstrings ordered consistently as the binary vector $z$.

Energy estimation is performed from measured histograms. The function
\begin{verbatim}
estimate_energy_from_histogram(histogram, Q, q_linear)
\end{verbatim}
computes the mean sampled QUBO energy. The function
\begin{verbatim}
estimate_cvar_energy_from_histogram(histogram, Q, q_linear, alpha)
\end{verbatim}
computes the CVaR energy by sorting sampled bitstrings according to their QUBO costs and averaging only the best fraction specified by \texttt{alpha}. The CVaR mode emphasizes low cost bitstrings instead of averaging over the entire sampled distribution.

The main internal component is the cached circuit evaluator. This component stores the circuit infrastructure that remains fixed during training. It stores the execution mode, QUBO data, ansatz depth, QPU topology, parameterized circuit, measured circuit, transpiled circuit template when available, and parameter order. For each new parameter vector, the evaluator binds the values to the cached circuit, executes the software, obtains the measured histogram, and returns the sampled energy.

The warm start functions implement the three metaheuristic initialization options. The functions
\begin{verbatim}
black_hole_optimize_vqe(...)
gwo_optimize_vqe(...)
abc_optimize_vqe(...)
\end{verbatim}
all use the cached evaluator to score candidate ansatz parameter vectors. Each candidate is evaluated by executing the variational circuit and estimating its sampled energy. The function
\begin{verbatim}
initialize_parameters(...)
\end{verbatim}
selects the requested initialization method. It returns a random parameter vector when random initialization is requested, or the best parameter vector found by the selected metaheuristic method.

The refinement stage is implemented using a simultaneous perturbation optimizer. The class
\begin{verbatim}
SPSAOptimizer
\end{verbatim}
generates two perturbed parameter vectors per iteration and estimates a descent direction from the difference between their sampled energies. This choice keeps the number of sampled energy evaluations per iteration independent of the number of ansatz parameters.

After training, the final solution is obtained from the sampled histogram. The function
\begin{verbatim}
best_solution_from_histogram(histogram, Q, q_linear)
\end{verbatim}
evaluates the original QUBO cost for the sampled bitstrings and returns the lowest cost bitstring found. This is different from returning the most frequent bitstring, because the most frequent bitstring is not always the best QUBO solution.

The outputs of \texttt{dvqe} are
\begin{verbatim}
z_best, final_circuit, hist
\end{verbatim}
where \texttt{z\_best} is the best sampled binary vector according to the original QUBO objective, \texttt{final\_circuit} is the trained variational circuit without the measurement wrapper, and \texttt{hist} is the final measurement histogram. The histogram can be used to inspect the sampled distribution, compare candidate bitstrings, and evaluate the probability of observing low cost solutions.

Overall, the DVQE software package follows the same logic as the algorithm in Subsection~\ref{subsec:proposed_dvqe}, but implements it through a cached and modular code structure. The package builds the circuit infrastructure once, evaluates ansatz parameters through sampled QUBO energy, supports random and metaheuristic warm starts, refines the parameters through simultaneous perturbation, and decodes the final answer by evaluating the original QUBO cost over the final sampled bitstrings.

\subsection{Proposed Sequential QP Reformulation as QUBO for DVQE Compatibility}
\label{subsec:qqp_algorithm}

The proposed QQP framework extends QUBO-based optimization from binary variables to bounded constrained QP problems. The main idea is to avoid a direct multi-bit discretization of each continuous variable. Instead, the proposed framework uses a nested sequential construction with three levels. The outer level scales the original bounded QP into a normalized variable space. The middle level handles equality and inequality constraints through a Powell-Hestenes-Rockafellar (PHR) augmented-Lagrangian fixed-region construction. The inner level solves each resulting bounded QP by repeatedly constructing and solving local one-bit QUBO subproblems. Algorithm~\ref{alg:qqp_algorithm} summarizes the proposed sequential QP to QUBO construction.

\begin{algorithm}[H]
\small
\caption{Proposed Sequential QP Reformulation as QUBO }
\label{alg:qqp_algorithm}
\DontPrintSemicolon
\SetAlgoLined
\SetKwInOut{Input}{Input}
\SetKwInOut{Output}{Output}

\Input{QP objective, linear equality constraints, linear inequality constraints, variable bounds, QUBO solver}
\Output{Approximate continuous solution $x^\star$}

Scale the original variable by writing $x=x_{\mathrm{c}}+D_xy$ with $-\mathbf{1}\leq y\leq \mathbf{1}$\;

Transform the QP objective and all constraints from $x$-space to $y$-space\;

Normalize constraint rows and scale the objective coefficients\;

Initialize the PHR augmented-Lagrangian multipliers and penalty parameter\;

\For{each augmented-Lagrangian iteration}{
Identify the current PHR region of the inequality constraints\;

Construct the proposed fixed-region quadratic model in $y$-space\;

Solve the fixed-region bounded QP using sequential one-bit QUBO refinement\;

Recompute the PHR region at the new point\;

Repeat the fixed-region construction until the region is stable or the region iteration limit is reached\;

Update the equality and inequality multipliers\;

Update the penalty parameter when needed\;

Stop if feasibility and stationarity tolerances are satisfied\;
}

Map the final normalized solution back to the original variable space\;

\Return{$x^\star$}
\end{algorithm}

The starting point is the bounded constrained QP
\begin{equation}
\begin{aligned}
\min_{x\in\mathbb{R}^{n}} \quad
& f(x)=x^T A x+b^T x+c \\
\mathrm{s.t.}\quad
& Gx=r,\\
& Hx\leq h,\\
& \ell\leq x\leq u .
\end{aligned}
\label{eq:qqp_original_qp}
\end{equation}
Here, $x$ is the continuous decision vector, $A$ and $b$ define the quadratic objective, $Gx=r$ represents equality constraints, $Hx\leq h$ represents inequality constraints, and $\ell\leq x\leq u$ gives the variable bounds. The proposed QQP framework transforms this constrained continuous QP into a sequence of local QUBO subproblems that can be solved by a QUBO backend.

The first step in the proposed QQP construction is variable normalization. The original box $\ell\leq x\leq u$ is mapped to the normalized box $-\mathbf{1}\leq y\leq \mathbf{1}$ by defining
\begin{equation}
x_{\mathrm{c}}=\frac{\ell+u}{2},
\qquad
D_x=\mathrm{diag}\left(\frac{u-\ell}{2}\right),
\qquad
x=x_{\mathrm{c}}+D_x y .
\label{eq:qqp_x_scaling}
\end{equation}
This transformation is useful because the original variables may have different units and ranges. After the transformation, every normalized variable lies in the same interval. This makes the later trust-region and one-bit QUBO steps more balanced.

Substituting \eqref{eq:qqp_x_scaling} into the objective gives a quadratic function in $y$:
\begin{equation}
f_y(y)=y^T A_y y+b_y^T y+c_y ,
\label{eq:qqp_y_objective}
\end{equation}
where
\begin{equation}
A_y=D_x^T A D_x,
\label{eq:qqp_Ay}
\end{equation}
\begin{equation}
b_y=2D_x^T A x_{\mathrm{c}}+D_x^T b,
\label{eq:qqp_by}
\end{equation}
and
\begin{equation}
c_y=x_{\mathrm{c}}^T A x_{\mathrm{c}}+b^T x_{\mathrm{c}}+c .
\label{eq:qqp_cy}
\end{equation}

The linear constraints are transformed in the same way. The equality constraints become
\begin{equation}
G_y y=r_y,
\label{eq:qqp_scaled_eq}
\end{equation}
where
\begin{equation}
G_y=GD_x,
\qquad
r_y=r-Gx_{\mathrm{c}}.
\label{eq:qqp_Gy_ry}
\end{equation}
The inequality constraints become
\begin{equation}
H_y y\leq h_y,
\label{eq:qqp_scaled_ineq}
\end{equation}
where
\begin{equation}
H_y=HD_x,
\qquad
h_y=h-Hx_{\mathrm{c}}.
\label{eq:qqp_Hy_hy}
\end{equation}
Thus, the scaled constrained QP is
\begin{equation}
\begin{aligned}
\min_{y\in\mathbb{R}^{n}} \quad
& y^T A_y y+b_y^T y+c_y \\
\mathrm{s.t.}\quad
& G_y y=r_y,\\
& H_y y\leq h_y,\\
& -\mathbf{1}\leq y\leq \mathbf{1}.
\end{aligned}
\label{eq:qqp_scaled_qp}
\end{equation}

The proposed implementation also improves numerical scaling before applying the augmented-Lagrangian construction. Each nonzero row of $G_y$ and $H_y$ can be divided by its Euclidean norm. For an equality row $g_i^T y=r_i$, the normalized row is
\begin{equation}
\bar{g}_i^T y=\bar{r}_i,
\qquad
\bar{g}_i=\frac{g_i}{\|g_i\|_2},
\qquad
\bar{r}_i=\frac{r_i}{\|g_i\|_2}.
\label{eq:qqp_row_scaled_eq}
\end{equation}
For an inequality row $h_i^T y\leq \eta_i$, the normalized row is
\begin{equation}
\bar{h}_i^T y\leq \bar{\eta}_i,
\qquad
\bar{h}_i=\frac{h_i}{\|h_i\|_2},
\qquad
\bar{\eta}_i=\frac{\eta_i}{\|h_i\|_2}.
\label{eq:qqp_row_scaled_ineq}
\end{equation}
This row normalization does not change the feasible set, but it prevents constraints with large coefficients from dominating the augmented-Lagrangian terms only because of numerical scale. The objective coefficients may also be divided by a positive scalar. Since multiplication of the whole objective by a positive constant does not change its minimizer, this scaling improves numerical conditioning without changing the solution.

After scaling, the proposed QQP framework handles the constrained QP using a PHR augmented Lagrangian. Define the equality and inequality residuals
\begin{equation}
e(y)=G_y y-r_y,
\qquad
g(y)=H_y y-h_y.
\label{eq:qqp_residuals}
\end{equation}
For equality multipliers $\lambda_e$, inequality multipliers $\lambda_g\geq 0$, and penalty parameter $\mu>0$, the PHR augmented Lagrangian is
\begin{equation}
\begin{aligned}
\mathcal{L}_{\mu}(y,\lambda_e,\lambda_g)
=
& f_y(y)
+\lambda_e^T e(y)
+\frac{\mu}{2}\|e(y)\|^2 \\
&+
\frac{1}{2\mu}
\left(
\left\|
\max\left(0,\lambda_g+\mu g(y)\right)
\right\|^2
-
\|\lambda_g\|^2
\right).
\end{aligned}
\label{eq:qqp_phr_al}
\end{equation}
The maximum is applied componentwise. This form handles inequality constraints directly and does not introduce slack variables. Avoiding slack variables is important in the proposed QUBO-based framework because every additional continuous variable would increase the dimension of the local QUBO subproblems.

The inequality part of \eqref{eq:qqp_phr_al} is piecewise quadratic. At a given point, define the PHR region
\begin{equation}
\mathcal{P}
=
\left\{
i:
\lambda_{g,i}+\mu g_i(y)>0
\right\}.
\label{eq:qqp_phr_region}
\end{equation}
For indices $i\in\mathcal{P}$, the maximum term is active. For indices $i\notin\mathcal{P}$, the maximum term contributes no quadratic term in $y$. Therefore, if the region $\mathcal{P}$ is fixed, the PHR augmented Lagrangian becomes an ordinary quadratic function of $y$.

The proposed fixed-region quadratic model can be written explicitly. Let $H_{\mathcal{P}}$ be the submatrix of $H_y$ containing only the rows indexed by $\mathcal{P}$, let $h_{\mathcal{P}}$ be the corresponding right-hand side, and let $\lambda_{\mathcal{P}}$ be the corresponding inequality multipliers. For a fixed region $\mathcal{P}$, the augmented objective has the form
\begin{equation}
\mathcal{L}_{\mu}^{\mathcal{P}}(y)
=
y^T A_{\mathcal{P}} y
+
b_{\mathcal{P}}^T y
+
c_{\mathcal{P}} ,
\label{eq:qqp_fixed_region_model}
\end{equation}
where
\begin{equation}
A_{\mathcal{P}}
=
A_y
+
\frac{\mu}{2}G_y^T G_y
+
\frac{\mu}{2}H_{\mathcal{P}}^T H_{\mathcal{P}},
\label{eq:qqp_fixed_region_A}
\end{equation}
\begin{equation}
b_{\mathcal{P}}
=
b_y
+
G_y^T\lambda_e
-
\mu G_y^T r_y
+
H_{\mathcal{P}}^T\lambda_{\mathcal{P}}
-
\mu H_{\mathcal{P}}^T h_{\mathcal{P}},
\label{eq:qqp_fixed_region_b}
\end{equation}
and
\begin{equation}
\begin{aligned}
c_{\mathcal{P}}
=
&c_y
-\lambda_e^T r_y
+\frac{\mu}{2}r_y^T r_y
-\lambda_{\mathcal{P}}^T h_{\mathcal{P}}
+\frac{\mu}{2}h_{\mathcal{P}}^T h_{\mathcal{P}} \\
&-\frac{1}{2\mu}\lambda_g^T\lambda_g .
\end{aligned}
\label{eq:qqp_fixed_region_c}
\end{equation}
The constant term does not affect the minimizer of the fixed-region QP, but it makes the fixed-region objective value consistent with the PHR expression. For a fixed $\mathcal{P}$, the proposed method solves
\begin{equation}
\begin{aligned}
\min_{y\in\mathbb{R}^{n}} \quad
& y^T A_{\mathcal{P}} y+b_{\mathcal{P}}^T y+c_{\mathcal{P}} \\
\mathrm{s.t.}\quad
& -\mathbf{1}\leq y\leq \mathbf{1}.
\end{aligned}
\label{eq:qqp_fixed_region_qp}
\end{equation}
This is a bounded QP without explicit equality or inequality constraints. The original constraints are represented through the fixed-region PHR augmented objective.

The bounded QP in \eqref{eq:qqp_fixed_region_qp} is solved by the proposed inner sequential one-bit QUBO procedure. Consider a general bounded quadratic problem
\begin{equation}
\begin{aligned}
\min_{y\in\mathbb{R}^{n}} \quad
& \phi(y)=y^T A_r y+b_r^T y+c_r\\
\mathrm{s.t.}\quad
& \ell_y\leq y\leq u_y ,
\end{aligned}
\label{eq:qqp_inner_bounded_qp}
\end{equation}
where $A_r$, $b_r$, and $c_r$ come from the fixed-region model. At a current point $y^k$, the proposed method introduces a normalized local variable $\alpha$ by
\begin{equation}
y=y^k+D_r\alpha,
\qquad
D_r=\mathrm{diag}(u_y-\ell_y).
\label{eq:qqp_alpha_mapping}
\end{equation}
The vector $\alpha$ is restricted by both the variable bounds and a trust-region radius $\rho$:
\begin{equation}
\alpha_i^{\min}
=
\max\left\{
-\rho,
\frac{\ell_{y,i}-y_i^k}{u_{y,i}-\ell_{y,i}}
\right\},
\qquad
\alpha_i^{\max}
=
\min\left\{
\rho,
\frac{u_{y,i}-y_i^k}{u_{y,i}-\ell_{y,i}}
\right\}.
\label{eq:qqp_alpha_bounds}
\end{equation}
This step ensures that every point generated from $\alpha$ remains inside the bounds. The trust-region radius $\rho$ prevents the local QUBO from making a move that is too large for the current sequential refinement stage.

Substituting \eqref{eq:qqp_alpha_mapping} into \eqref{eq:qqp_inner_bounded_qp} gives an alpha-space quadratic problem
\begin{equation}
\phi_{\alpha}(\alpha)
=
\alpha^T A_{\alpha}\alpha
+
b_{\alpha}^T\alpha
+
c_{\alpha},
\label{eq:qqp_alpha_qp}
\end{equation}
with
\begin{equation}
A_{\alpha}=D_r^T A_r D_r,
\label{eq:qqp_A_alpha}
\end{equation}
\begin{equation}
b_{\alpha}=2D_r^T A_r y^k+D_r^T b_r,
\label{eq:qqp_b_alpha}
\end{equation}
and
\begin{equation}
c_{\alpha}=(y^k)^T A_r y^k+b_r^T y^k+c_r .
\label{eq:qqp_c_alpha}
\end{equation}
Thus, solving the bounded QP locally is equivalent to improving this alpha-space quadratic over the bounded interval $\alpha^{\min}\leq \alpha\leq \alpha^{\max}$.

The QUBO step is performed locally inside alpha-space. Let $\alpha^j$ be the current alpha iterate inside the inner refinement loop. For each component, the proposed method defines a positive and negative reachable distance:
\begin{equation}
d_i^{+}
=
\min\left\{
\delta_i,
\alpha_i^{\max}-\alpha_i^j
\right\},
\qquad
d_i^{-}
=
\min\left\{
\delta_i,
\alpha_i^j-\alpha_i^{\min}
\right\},
\label{eq:qqp_dplus_dminus}
\end{equation}
where $\delta_i$ is the current local step size. The lower local endpoint is
\begin{equation}
s_i=\alpha_i^j-d_i^{-},
\label{eq:qqp_local_endpoint}
\end{equation}
and the local binary step width is
\begin{equation}
p_i=d_i^{+}+d_i^{-}.
\label{eq:qqp_local_width}
\end{equation}
Using these quantities, a binary vector $z\in\{0,1\}^{n}$ represents the local alpha candidate
\begin{equation}
\alpha_{\mathrm{cand}}=s+Pz,
\qquad
P=\mathrm{diag}(p).
\label{eq:qqp_alpha_binary_map}
\end{equation}
If $z_i=0$, the $i$th alpha component takes its lower local endpoint $s_i$. If $z_i=1$, it takes the upper local endpoint $s_i+p_i$. Therefore, each continuous variable is represented by one local binary choice at the current refinement scale.

Substituting \eqref{eq:qqp_alpha_binary_map} into \eqref{eq:qqp_alpha_qp} gives
\begin{equation}
\begin{aligned}
\phi_{\alpha}(s+Pz)
=
& (s+Pz)^T A_{\alpha}(s+Pz)
+
b_{\alpha}^T(s+Pz)
+
c_{\alpha} \\
=
& z^T Q_{\mathrm{loc}} z
+
q_{\mathrm{loc}}^T z
+
\kappa_{\mathrm{loc}},
\end{aligned}
\label{eq:qqp_local_qubo_expansion}
\end{equation}
where
\begin{equation}
Q_{\mathrm{loc}}=P^T A_{\alpha} P,
\label{eq:qqp_Q_local}
\end{equation}
\begin{equation}
q_{\mathrm{loc}}
=
2P^T A_{\alpha}s+P^T b_{\alpha},
\label{eq:qqp_q_local}
\end{equation}
and
\begin{equation}
\kappa_{\mathrm{loc}}
=
s^T A_{\alpha}s+b_{\alpha}^T s+c_{\alpha}.
\label{eq:qqp_kappa_local}
\end{equation}
The constant $\kappa_{\mathrm{loc}}$ does not change the binary minimizer. Therefore, the continuous local search step is reduced to the QUBO
\begin{equation}
\min_{z\in\{0,1\}^{n}}
\quad
z^T Q_{\mathrm{loc}} z+q_{\mathrm{loc}}^T z .
\label{eq:qqp_local_qubo}
\end{equation}
This is the main compatibility step between the proposed continuous QP reformulation and a QUBO solver.

The dimension of \eqref{eq:qqp_local_qubo} is $n$, the number of original continuous variables. It does not depend on a multi-bit discretization level. Therefore, the proposed method avoids the common expansion in which each continuous variable is represented by several binary bits. Accuracy is instead obtained through repeated local refinements. Large steps provide exploration early in the process, while smaller steps provide local accuracy later.

After the local QUBO is solved, the binary solution $z^\star$ is decoded into an alpha candidate:
\begin{equation}
\alpha_{\mathrm{cand}}=s+Pz^\star .
\label{eq:qqp_decode_alpha}
\end{equation}
The corresponding objective value is evaluated using the alpha-space quadratic:
\begin{equation}
\phi_{\alpha}(\alpha_{\mathrm{cand}})
=
\alpha_{\mathrm{cand}}^T A_{\alpha}\alpha_{\mathrm{cand}}
+
b_{\alpha}^T\alpha_{\mathrm{cand}}
+
c_{\alpha}.
\label{eq:qqp_alpha_candidate_obj}
\end{equation}
The candidate is accepted only if
\begin{equation}
\phi_{\alpha}(\alpha^j)
-
\phi_{\alpha}(\alpha_{\mathrm{cand}})
>
\tau_f,
\label{eq:qqp_inner_acceptance}
\end{equation}
where $\tau_f$ is the objective improvement tolerance. If this condition holds, the alpha iterate is updated:
\begin{equation}
\alpha^{j+1}=\alpha_{\mathrm{cand}}.
\label{eq:qqp_alpha_accept}
\end{equation}
If the condition does not hold, the candidate is rejected and the local step size is reduced:
\begin{equation}
\delta^{j+1}=\gamma_{\alpha}\delta^j,
\qquad
0<\gamma_{\alpha}<1.
\label{eq:qqp_alpha_step_reduce}
\end{equation}
This accept-reject mechanism is important because the QUBO solver returns the best binary point for the current local two-point representation, but that point may not improve the continuous quadratic model. Rejection and step reduction allow the proposed local search to refine the resolution without forcing a non-improving move.

The inner one-bit QUBO refinement continues until the relative local step size becomes small, the maximum number of inner refinements is reached, or repeated rejected QUBO proposals indicate that further refinement in the current local region is not useful. After the inner loop terminates, the final alpha vector is mapped back to the bounded QP variable:
\begin{equation}
y_{\mathrm{new}}=y^k+D_r\alpha .
\label{eq:qqp_alpha_to_y}
\end{equation}
This point is the approximate solution of the current fixed-region bounded QP.

The proposed fixed-region loop then checks whether the PHR region used to build the bounded QP is consistent with the new point. The region is recomputed as
\begin{equation}
\mathcal{P}_{\mathrm{new}}
=
\left\{
i:
\lambda_{g,i}+\mu g_i(y_{\mathrm{new}})>0
\right\}.
\label{eq:qqp_new_region}
\end{equation}
If $\mathcal{P}_{\mathrm{new}}=\mathcal{P}$, the fixed-region model is stable, and the augmented-Lagrangian iteration can proceed. If the region changes, a new fixed-region quadratic model is constructed using $\mathcal{P}_{\mathrm{new}}$, and the bounded QP solve is repeated. This step is necessary because the PHR objective is piecewise quadratic. A quadratic model built for one region may no longer represent the correct PHR expression after the point moves into another region.

After the fixed-region loop terminates, the augmented-Lagrangian multipliers are updated. Equality multipliers are updated by
\begin{equation}
\lambda_e^{k+1}
=
\lambda_e^k+\mu e(y^{k+1}),
\label{eq:qqp_lambda_eq_update}
\end{equation}
and inequality multipliers are updated by the projected rule
\begin{equation}
\lambda_g^{k+1}
=
\max\left(0,\lambda_g^k+\mu g(y^{k+1})\right).
\label{eq:qqp_lambda_ineq_update}
\end{equation}
The projection is required because inequality multipliers must remain nonnegative. The penalty parameter may also be adjusted. If constraint violation is not decreasing sufficiently, $\mu$ is increased to enforce feasibility more strongly. If the dual progress dominates the primal progress, $\mu$ may be decreased to avoid an overly stiff penalty term.

The augmented-Lagrangian loop stops when the equality residuals, inequality violations, and stationarity measure satisfy the prescribed tolerances, or when the maximum iteration budget is reached. The final solution in normalized variables is denoted by $y^\star$. The solution is then mapped back to the original variable space:
\begin{equation}
x^\star=x_{\mathrm{c}}+D_x y^\star.
\label{eq:qqp_final_x}
\end{equation}
The final objective and feasibility residuals are evaluated in the original $x$-space.

Overall, the proposed QQP framework makes bounded constrained QP problems compatible with QUBO solvers through a nested sequential construction. First, the original QP is scaled into a normalized bounded form. Second, the constrained QP is handled by fixed-region PHR augmented-Lagrangian models without adding slack variables. Third, each bounded QP is solved through repeated one-bit QUBO refinements. Each QUBO proposes one local binary decision per continuous variable, and the decoded result becomes a continuous step only if it improves the local model. Although the presentation above focuses on QP problems, bounded LP problems can also be handled as the special case obtained by setting the quadratic objective matrix to zero. This structure allows a QUBO solver such as DVQE to serve as the binary optimization engine inside the proposed continuous optimization framework.

\subsection{QQP Software Package}
\label{subsec:qqp_package}

The QQP framework is implemented as a Python software package for solving bounded QP and LP problems by using QUBO solvers as the inner optimization engine. The package is designed to connect continuous quadratic or linear optimization models to quantum-compatible QUBO subproblems. It provides a high level function for the user, while internally separating scaling, augmented-Lagrangian handling, fixed-region QP construction, sequential one-bit QUBO refinement, and final solution recovery.

The main entry point of the package is the function
\begin{verbatim}
qqp(...)
\end{verbatim}
which receives the original QP or LP data and returns an approximate continuous solution. The QUBO subproblems can be solved by the DVQE software package or by a classical MIQP solver for comparison. A typical call has the form
\begin{verbatim}
x_sol, f_original, al_iters, info = qqp(
    A=A,
    b=b,
    c=c,
    G=G,
    r=r,
    H=H,
    h=h,
    x0=x0,
    lb=lb,
    ub=ub,
    qubo_solver="dvqe",
    mode="distributed",
    init_type=2,
    depth=1,
    lr=0.08,
    max_iters=30,
    qpu_qubit_config=[4, 3, 2],
    num_shots=128,
    final_shots=1000,
    warm_start_population=4,
    warm_start_iters=5,
    warm_start_shots=64,
    energy_mode="cvar",
    cvar_alpha=0.2
)
\end{verbatim}

The inputs \texttt{A}, \texttt{b}, and \texttt{c} define the objective
\[
f(x)=x^T A x+b^T x+c .
\]
For an LP, the matrix \texttt{A} can be set to the zero matrix, so the same interface is used for both QP and LP problems. The inputs \texttt{G} and \texttt{r} define equality constraints while \texttt{H} and \texttt{h} define inequality constraints. The vectors \texttt{lb} and \texttt{ub} define the lower and upper bounds on the continuous variables. The optional input \texttt{x0} gives the initial point. If no initial point is provided, the package starts from the center of the box constraints. The first part of \texttt{qqp(...)} performs variable scaling. The original variable is transformed to $y$-space. The scaling controls are
\begin{verbatim}
normalize_constraints=True
scale_objective=True
regularize_weak_cost=True
eps_reg=1e-8
\end{verbatim}
The option \texttt{normalize\_constraints} normalizes each nonzero constraint row by its Euclidean norm. The option \texttt{scale\_objective} divides the objective coefficients by a positive scale factor so that the QP coefficients are better conditioned. The option \texttt{regularize\_weak\_cost} adds a small regularization term to improve variables that are weakly represented in the objective. The parameter \texttt{eps\_reg} controls the size of this regularization.

After scaling, \texttt{qqp(...)} calls
\begin{verbatim}
dqp(...)
\end{verbatim}
to solve the scaled constrained QP. The function \texttt{dqp(...)} handles equality and inequality constraints using a fixed-region PHR augmented-Lagrangian method. Its inputs include the scaled objective, scaled equality constraints, scaled inequality constraints, normalized bounds, initial point, penalty parameters, multiplier settings, stopping tolerances, and the settings passed to the inner QUBO-based solver.

The main augmented-Lagrangian controls include
\begin{verbatim}
mu0
mu_min
mu_max
constraint_tol
stationarity_tol
max_al_iters
min_al_iters
update_mu
mu_update_rule
\end{verbatim}
The parameter \texttt{mu0} is the initial penalty parameter. The values \texttt{mu\_min} and \texttt{mu\_max} define lower and upper limits for penalty updates. The parameter \texttt{constraint\_tol} controls the feasibility tolerance, while \texttt{stationarity\_tol} can be used to enforce a stationarity check. The parameters \texttt{max\_al\_iters} and \texttt{min\_al\_iters} control the augmented-Lagrangian iteration range. The option \texttt{update\_mu} activates penalty updates, and \texttt{mu\_update\_rule} selects the update logic.

Inside \texttt{dqp(...)}, the PHR inequality term is treated through fixed-region construction. The function identifies the current PHR region based on the shifted inequality residuals. For the selected region, the augmented-Lagrangian expression becomes an ordinary quadratic function. The package constructs the corresponding fixed-region QP through the helper
\begin{verbatim}
_build_phr_fixed_region_qp(...)
\end{verbatim}
This helper returns the fixed-region quadratic objective coefficients. It includes the contribution of the original objective, equality penalty terms, active PHR inequality terms, multiplier terms, and the constant terms needed to keep the fixed-region objective consistent with the PHR expression.

The fixed-region settings are
\begin{verbatim}
phr_region_tol
max_region_iters
\end{verbatim}
The parameter \texttt{phr\_region\_tol} controls the tolerance used to decide whether an inequality belongs to the active PHR region. The parameter \texttt{max\_region\_iters} limits how many times the package rebuilds and resolves a fixed-region QP inside one augmented-Lagrangian iteration. This is needed because the solution of one fixed-region QP may move to another PHR region.

Each fixed-region bounded QP is solved by
\begin{verbatim}
dqup(...)
\end{verbatim}
This function is the sequential one-bit QUBO solver for bounded continuous quadratic problems. It receives a bounded QP of the form
\[
\min_x x^T A x+b^T x+c,
\qquad
lb\leq x\leq ub,
\]
and solves it approximately by repeatedly constructing local QUBO subproblems. The variable is first mapped into a local alpha space around the current point. This construction is performed by
\begin{verbatim}
_build_alpha_qup(...)
\end{verbatim}
which returns the alpha-space quadratic coefficients, the alpha bounds, and the variable range used to decode the solution back to the original bounded QP variable.

The inner one-bit QUBO is constructed by
\begin{verbatim}
_build_local_alpha_qubo(...)
\end{verbatim}
This helper creates a local binary representation of the alpha variable. For each continuous variable, it defines a lower local endpoint and a local binary step width. The binary vector chooses between the two local endpoints for each component. Substituting this local binary representation into the alpha-space quadratic gives one QUBO with the same dimension as the number of continuous variables. The constant term is stored for objective accounting, but the QUBO solver only needs the quadratic and linear QUBO coefficients.

The local QUBO solver is selected by the input
\begin{verbatim}
qubo_solver
\end{verbatim}
When \texttt{qubo\_solver="dvqe"}, the local QUBO is sent to the DVQE software package. The DVQE-related inputs, such as \texttt{mode}, \texttt{init\_type}, \texttt{depth}, \texttt{lr}, \texttt{max\_iters}, \texttt{qpu\_qubit\_config}, \texttt{num\_shots}, \texttt{final\_shots}, \texttt{warm\_start\_population}, \texttt{warm\_start\_iters}, \texttt{warm\_start\_shots}, \texttt{energy\_mode}, and \texttt{cvar\_alpha}, are passed to the DVQE call. When \texttt{qubo\_solver="classical"}, the local QUBO is solved as a classical binary MIQP using Gurobi. This option is useful for benchmarking and debugging.

Before a local QUBO is sent to DVQE, its coefficients can be scaled by
\begin{verbatim}
_scale_qubo(...)
\end{verbatim}
This scaling improves numerical stability for variational QUBO training. Since the QUBO objective is divided by a positive constant, the minimizer does not change. Classical MIQP solving can use the original local QUBO coefficients directly.

After the QUBO solver returns a binary vector, the helper
\begin{verbatim}
_decode_alpha_solution(...)
\end{verbatim}
maps the binary solution back to a continuous alpha candidate. The candidate is then evaluated in the alpha-space quadratic model. If it improves the local objective by more than the specified tolerance, the candidate is accepted. If it does not improve the objective, the local alpha step is reduced and another local QUBO is built. This accept-reject structure prevents the continuous solver from accepting non-improving QUBO proposals.

The main controls for the bounded QP refinement in \texttt{dqup(...)} include
\begin{verbatim}
rho0
rho_decay
rho_tol
obj_tol
max_outer_iters
alpha_step0
alpha_step_fraction
alpha_step_decay
alpha_step_tol
max_inner_iters
alpha_warm_start
alpha_warm_start_scale
\end{verbatim}
The parameter \texttt{rho0} sets the initial trust-region radius in the outer continuous search. The parameter \texttt{rho\_decay} shrinks this radius when outer updates are rejected. The value \texttt{rho\_tol} stops the outer loop if the region becomes too small. The parameter \texttt{obj\_tol} defines the minimum objective improvement needed to accept a candidate. The parameters \texttt{alpha\_step0}, \texttt{alpha\_step\_fraction}, \texttt{alpha\_step\_decay}, and \texttt{alpha\_step\_tol} control the inner alpha step size. The option \texttt{alpha\_warm\_start} allows the previous alpha solution to initialize the next local problem, and \texttt{alpha\_warm\_start\_scale} controls how strongly this previous alpha step is reused.

The early stopping controls are
\begin{verbatim}
max_consecutive_inner_rejections
max_consecutive_outer_rejections
\end{verbatim}
The first parameter stops the inner alpha loop after repeated QUBO candidates fail to improve the local alpha objective. The second parameter stops the outer continuous loop after repeated outer candidates fail to improve the bounded QP objective. These controls reduce unnecessary QUBO calls when the current region is no longer productive.

The output of \texttt{dqup(...)} is
\begin{verbatim}
x_best, f_best, total_outer_iterations, info
\end{verbatim}
when \texttt{return\_info=True}. Here, \texttt{x\_best} is the best bounded QP solution found by the sequential one-bit QUBO process, \texttt{f\_best} is its objective value, and \texttt{total\_outer\_iterations} is the number of completed outer iterations. The dictionary \texttt{info} stores the trajectory, objective history, trust-region history, alpha histories, local QUBO histories when requested, QUBO solver information, runtime data, and acceptance status.

The output of \texttt{dqp(...)} is
\begin{verbatim}
x, f_final_original, total_al_iterations, info
\end{verbatim}
when \texttt{return\_info=True}. Here, \texttt{x} is the final solution of the constrained scaled QP, \texttt{f\_final\_original} is the objective value in the scaled problem handled by \texttt{dqp(...)}, and \texttt{total\_al\_iterations} is the number of augmented-Lagrangian iterations. The dictionary \texttt{info} includes multiplier histories, penalty histories, residual histories, fixed-region histories, DQUP histories, stationarity and complementarity diagnostics, and final feasibility information.

After \texttt{dqp(...)} returns the solution in normalized coordinates, \texttt{qqp(...)} maps it back to the original variable space. The final outputs of \texttt{qqp(...)} are
\begin{verbatim}
x_sol, f_original, al_iters, info
\end{verbatim}
when \texttt{return\_info=True}. The output \texttt{x\_sol} is the final solution in the original $x$-space. The output \texttt{f\_original} is the original objective value evaluated at \texttt{x\_sol}. The output \texttt{al\_iters} is the number of augmented-Lagrangian iterations. The dictionary \texttt{info} stores both the scaled problem data and the original-space recovery data, including the normalized solution, scaling information, original objective, final residuals, multiplier values, penalty values, and solver histories.

Overall, the QQP software package implements the algorithm in Subsection~\ref{subsec:qqp_algorithm} through a layered structure. The high level function \texttt{qqp(...)} scales the original QP or LP and recovers the final solution in the original space. The constrained solver \texttt{dqp(...)} handles linear equality and inequality constraints using fixed-region PHR augmented-Lagrangian QPs. The bounded solver \texttt{dqup(...)} converts each bounded QP into a sequence of one-bit QUBO subproblems. The QUBO subproblems are then solved by the DVQE software package or by a classical MIQP solver. This structure makes the package suitable for testing how quantum QUBO solvers can be used as components inside continuous QP and LP solution workflows.

\section{Experiments and Discussion}
\label{sec:experiments}

This section evaluates the proposed DVQE and QQP software packages. The DVQE experiments study the effect of execution mode and initialization strategy on QUBO solution quality, while the QQP experiments test the use of DVQE as a QUBO backend for bounded QP and LP problems. The software packages and all runner codes used in the experiments are available on \href{https://github.com/LSU-RAISE-LAB/DVQE.git}{GitHub} \cite{GitHub_raiselab}.

\subsubsection{Ablation Study on DVQE}
\label{subsubsec:dvqe_ablation}

The first experiment evaluates the effect of execution mode and initialization strategy on the proposed DVQE solver. The goal is to test whether the distributed implementation preserves the solution quality of the monolithic implementation and whether the proposed metaheuristic warm-start stage improves the performance of DVQE compared with random initialization. This experiment is designed as an ablation study rather than as a maximum-accuracy tuning study. Therefore, the same moderate DVQE settings are used for all methods to provide an apple-to-apple comparison. Higher success rates could likely be obtained by increasing the ansatz depth, number of shots, warm-start population, warm-start iterations, or variational refinement budget. However, the purpose here is not to maximize the accuracy of every variant, but to compare the relative effect of execution mode and initialization under the same computational setting. Random QUBO instances were generated for problem sizes
\(
n=6,7,\ldots,15 .
\)
For each problem size, eight independent random seeds were used, giving a total of 80 QUBO instances. Each instance was solved using five DVQE variants:
\[
\begin{array}{ll}
\text{M-Rand:} & \text{monolithic DVQE with random initialization},\\
\text{D-Rand:} & \text{distributed DVQE with random initialization},\\
\text{D-BH:} & \text{distributed DVQE with Black Hole warm start},\\
\text{D-GWO:} & \text{distributed DVQE with Grey Wolf Optimizer warm start},\\
\text{D-ABC:} & \text{distributed DVQE with Artificial Bee Colony warm start}.
\end{array}
\]
Thus, the experiment contains 400 DVQE runs in total. The same QUBO instance was used for all five variants in each case. The exact optimum of each QUBO was computed by brute force and used as the reference value.

All DVQE training parameters were fixed across the five variants. The ansatz depth was set to 1, the learning rate was set to 0.05, and the maximum number of variational refinement iterations was set to 80. The relative stopping tolerance was $10^{-5}$. Each training energy evaluation used 256 shots, while the final sampling stage used 2000 shots. For warm-started variants, the population size was 6, the number of warm-start iterations was 8, and each warm-start evaluation used 64 shots. The sampled energy was evaluated using the CVaR estimator with $\alpha=0.2$. For distributed runs, the same QPU layout
\(
[4,4,4,4,4,4]
\)
was used in all cases. Therefore, any change in performance is due to either the execution mode or the initialization strategy, not to different training settings.

Table~\ref{tab:dvqe_ablation_summary} reports the aggregate results over all 80 QUBO instances. The monolithic random baseline solved 44 out of 80 instances exactly, giving a success rate of 55.0\%. The distributed random variant solved 42 out of 80 instances exactly, giving a success rate of 52.5\%. These two results are close, which indicates that the distributed implementation preserves the behavior of the monolithic DVQE solver under the same random initialization setting. The mean absolute errors are also close: 1.0954 for monolithic random and 1.1298 for distributed random.

The warm-started distributed variants improve the success rate compared with distributed random initialization. Among the three warm-start methods, the Black Hole initialization gives the best overall performance. The D-BH variant solves 66 out of 80 instances exactly, corresponding to a success rate of 82.5\%. It also gives the smallest mean absolute error, 0.2387. The D-ABC variant gives the second-best performance, with 60 successes and a success rate of 75.0\%. The D-GWO variant gives 52 successes and a success rate of 65.0\%, which is still better than distributed random initialization but weaker than D-BH and D-ABC in this experiment.

\begin{table}[!t]
\centering
\caption{DVQE ablation}
\label{tab:dvqe_ablation_summary}
\begin{tabular}{lcccccc}
\hline
Method & Runs & Successes & Success Rate & Mean Error & Max Error & Mean Time (s) \\
\hline
M-Rand & 80 & 44 & 0.550 & 1.0954 & 6.6274 & 1.0254 \\
D-Rand & 80 & 42 & 0.525 & 1.1298 & 11.6356 & 1.0317 \\
D-BH   & 80 & 66 & 0.825 & 0.2387 & 4.9005 & 1.2704 \\
D-GWO  & 80 & 52 & 0.650 & 0.5929 & 6.6398 & 1.2223 \\
D-ABC  & 80 & 60 & 0.750 & 0.3405 & 3.5388 & 1.6060 \\
\hline
\end{tabular}
\end{table}

The runtime results show that the warm-started methods require additional computational time because they perform a population-based search before the variational refinement stage. The two random-initialized variants have nearly the same average runtime, about 1.03 seconds per run. The D-BH and D-GWO variants increase the mean runtime to 1.2704 and 1.2223 seconds, respectively. The D-ABC variant is the most expensive among the tested methods, with an average runtime of 1.6060 seconds. Therefore, the warm-start stage improves solution quality at the cost of additional initialization time.

The best tradeoff in this experiment is obtained by the D-BH variant. It gives the highest success rate, the smallest mean absolute error, and a moderate runtime increase compared with distributed random initialization. The D-ABC variant is also effective, but it is slower and less accurate than D-BH on average. The D-GWO variant improves over random initialization, but its success rate and error are weaker than the other two warm-start methods.

These results support two main observations. First, the distributed DVQE implementation is consistent with the monolithic implementation under the same random-initialization setting. Second, metaheuristic warm starts can improve DVQE performance by guiding the variational circuit toward better regions of the parameter space before sampling-based refinement begins.

\subsubsection{Comparison Between D-BH and Gurobi MIQP}
\label{subsubsec:dbh_vs_gurobi}

The second DVQE experiment compares the best configuration selected from the ablation study against Gurobi MIQP. The purpose of this experiment is not to claim that the software-based D-BH implementation is faster than a mature classical MIQP solver. Instead, Gurobi is used as an exact reference solver, and D-BH is evaluated by how often it recovers the same optimal solution and how small its objective gap is when it does not. Hard random QUBO instances were generated for problem sizes
\(
n=3,4,\ldots,18 .
\)
For each problem size, five random seeds were used, resulting in 80 QUBO instances in total. Each QUBO was generated as a dense mixed-sign problem with many pairwise interactions, so that the resulting instances contain competing binary-variable effects and nontrivial local minima. The ansatz depth was set to 1, the learning rate was set to 0.5, and the maximum number of refinement iterations was set to 80. The relative stopping tolerance was $10^{-4}$. Each training energy evaluation used 512 shots, and the final sampling stage used 5000 shots. The Black Hole warm-start stage used a population size of 12, 20 warm-start iterations, and 512 shots per warm-start evaluation. The CVaR energy estimator was used with $\alpha=0.1$. The distributed QPU layout was fixed as
\(
[4,4,4,4,4]
\).

Table~\ref{tab:dbh_vs_gurobi_overall} summarizes the overall comparison. D-BH recovers the exact Gurobi objective and the same binary solution in 74 out of 80 instances, corresponding to a success rate of 92.5\%. The near-success rate is also 92.5\% under the tolerance $10^{-4}$. The mean absolute objective gap is 0.0687, while the median absolute gap is zero. This means that more than half of the tested instances are solved exactly, and the average error is mainly caused by a small number of harder cases. The mean relative gap is 0.0039, which indicates that even when D-BH does not exactly match Gurobi, the returned objective is usually close to the reference optimum.

\begin{table}[!t]
\centering
\caption{Comparison between Gurobi MIQP and D-BH}
\label{tab:dbh_vs_gurobi_overall}
\begin{tabular}{lc}
\hline
Metric & Value \\
\hline
Number of instances & 80 \\
D-BH exact success count & 74 \\
D-BH exact success rate & 0.925 \\
D-BH near success count & 74 \\
D-BH near success rate & 0.925 \\
D-BH same-solution count & 74 \\
D-BH same-solution rate & 0.925 \\
Mean absolute gap & 0.0687 \\
Median absolute gap & 0.0000 \\
Maximum absolute gap & 2.2673 \\
Mean relative gap & 0.0039 \\
Maximum relative gap & 0.0995 \\
Mean Hamming distance & 0.3750 \\
Maximum Hamming distance & 10 \\
Mean Gurobi runtime (s) & 0.0040 \\
Mean D-BH runtime (s) & 2.8434 \\
\hline
\end{tabular}
\end{table}

\begin{figure}[!t]
    \centering
    \includegraphics[width=\linewidth]{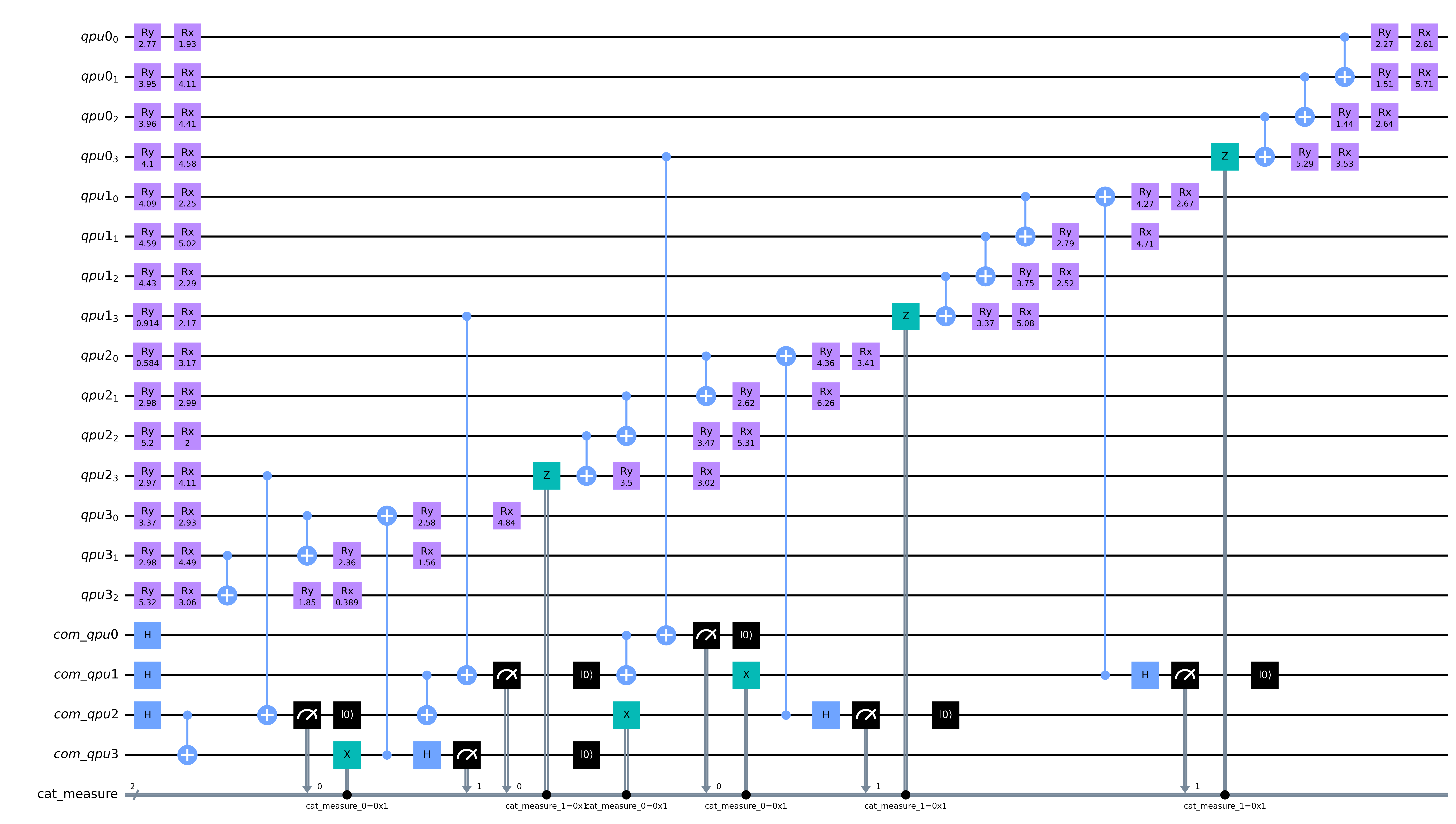}
    \caption{Trained distributed ansatz}
    \label{fig:trained_distributed_ansatz_n30}
\end{figure}

As expected, Gurobi is significantly faster in this experiment. Its mean runtime is 0.0040 seconds, while the mean D-BH runtime is 2.8434 seconds. This runtime difference is not surprising because the current D-BH implementation is evaluated through a CPU-based quantum-circuit software with repeated sampling, warm-start evaluation, and variational refinement. Therefore, the runtime comparison should not be interpreted as a claim of computational speedup. Rather, the important observation is that the proposed D-BH software can recover the same solution as Gurobi on most tested QUBO instances and returns small objective gaps when it fails to exactly match the MIQP optimum. A trained distributed circuit obtained after the D-BH workflow with depth 2 is shown in Fig.~\ref{fig:trained_distributed_ansatz_n30}.

\subsubsection{Validation of QQP}
\label{subsubsec:qqp_classical_validation}

This experiment validates the proposed QQP framework when DVQE is used as the internal QUBO solver. The purpose is to test whether constrained continuous QP and LP problems can be solved through the proposed two-level procedure: first, QQP reformulates the continuous constrained problem into a sequence of local one-bit QUBO subproblems; second, each local QUBO subproblem is solved using DVQE. Therefore, the continuous problem is not solved directly by DVQE. Instead, DVQE acts as the binary QUBO engine inside the QQP refinement loop with D-BH workflow. The test set contains both QP and LP instances with
\(
n=3,4,\ldots,32 .
\)
For QP instances, QQP is called with
\[
\texttt{qqp(..., qubo\_solver="dvqe")} ,
\]
while for LP instances, the quadratic matrix is set to zero and QQP is called with
\[
\texttt{qqp(..., A=0, qubo\_solver="dvqe")} .
\]
Each dimension includes one easy and one hard problem. The easy instances are well scaled and are used to test baseline accuracy. The hard instances are intentionally badly scaled, with scaled variables, scaled objective coefficients, dense constraints, and many active or nearly active inequalities. These hard cases are included to test whether the QQP--DVQE pipeline remains accurate under difficult numerical conditions.

This experiment uses percentage-based quality metrics. A run is counted as successful if the objective gap, normalized solution infinity-norm gap, and maximum normalized constraint gap are all below $1\%$, namely
\[
\max\{
\mathrm{ObjGap}_{\%},
\mathrm{xInfGap}_{\%},
\mathrm{ConstrGap}_{\%}
\}
\le 1\% .
\]
This criterion gives a scale-aware evaluation of QQP solution quality and avoids rejecting accurate solutions only because a badly scaled problem produces a large-looking absolute residual.

Table~\ref{tab:qqp_dvqe_validation} reports the percentage-based validation results for QQP with DVQE as the internal QUBO solver. The table focuses on solution quality rather than wall-clock time. The runtime of the present implementation is much larger than a direct classical solver because DVQE is executed through a CPU-based quantum-circuit software and because QQP may require many local QUBO calls during augmented-Lagrangian and trust-region refinement. Therefore, the timing results should be interpreted as software overhead, not as a claim of runtime advantage.

\begin{table}[!t]
\centering
\caption{Validation of QQP with DVQE as the internal QUBO solver}
\label{tab:qqp_dvqe_validation}
\tiny
\setlength{\tabcolsep}{2.5pt}
\renewcommand{\arraystretch}{1.15}
\begin{tabular}{
>{\raggedright\arraybackslash}p{0.30\linewidth}
>{\centering\arraybackslash}p{0.14\linewidth}
>{\centering\arraybackslash}p{0.14\linewidth}
>{\centering\arraybackslash}p{0.14\linewidth}
>{\centering\arraybackslash}p{0.14\linewidth}}
\toprule
Metric & Easy QP & Hard QP & Easy LP & Hard LP \\
\midrule
Cases & 30 & 30 & 30 & 30 \\
Reference optimal cases & 30 & 27 & 30 & 30 \\
QQP solved cases & 30 & 27 & 30 & 30 \\
$1\%$ quality successes & 30 & 26 & 24 & 30 \\
$1\%$ quality rate & 100.00\% & 96.30\% & 80.00\% & 100.00\% \\
Mean objective gap & 0.000238\% & 0.179209\% & 0.011012\% & 0.006548\% \\
Maximum objective gap & 0.001785\% & 4.708885\% & 0.084832\% & 0.029086\% \\
Mean solution infinity gap & 0.024851\% & 7.833827\% & 5.202572\% & 0.004404\% \\
Maximum solution infinity gap & 0.140097\% & 211.368821\% & 98.996008\% & 0.072967\% \\
Mean equality residual & 0.000557\% & 0.000432\% & 0.008254\% & 0.004236\% \\
Maximum equality residual & 0.002416\% & 0.003563\% & 0.078025\% & 0.073240\% \\
Mean inequality violation & 0.000160\% & $7.584{\times}10^{-5}$\% & 0.004406\% & 0.000902\% \\
Maximum inequality violation & 0.000544\% & 0.000280\% & 0.050943\% & 0.005664\% \\
Mean maximum constraint gap & 0.000558\% & 0.000444\% & 0.008320\% & 0.004490\% \\
Maximum constraint gap & 0.002416\% & 0.003563\% & 0.078025\% & 0.073240\% \\
DVQE time & High & High & High & High \\
\bottomrule
\end{tabular}
\end{table}

The QP results show strong performance. For easy QPs, all 30 cases satisfy the $1\%$ quality rule. The mean objective gap is only $0.000238\%$, and the maximum objective gap is only $0.001785\%$. The mean maximum constraint gap is also only $0.000558\%$. These values show that the QQP--DVQE pipeline accurately recovers the reference QP solutions on well-scaled constrained quadratic problems.

The hard QP results are more challenging because the problems are intentionally badly scaled. Nevertheless, 26 out of 27 reference-optimal cases satisfy the $1\%$ quality rule. The mean maximum constraint gap is only $0.000444\%$, and the maximum constraint gap is only $0.003563\%$. This shows that the augmented-Lagrangian, PHR fixed-region, and local QUBO refinement structure remains numerically stable when the QUBO layer is solved by DVQE. Some hard QP cases show large solution-vector percentage gaps, but these cases are dominated by scaling effects and sensitivity of the decision variables. The objective and constraint errors remain small in almost all cases.

The LP results also show accurate objective and feasibility behavior. For hard LPs, all 30 cases satisfy the $1\%$ quality rule. The mean objective gap is $0.006548\%$, the maximum objective gap is $0.029086\%$, and the maximum constraint gap is $0.073240\%$. These are all well below the $1\%$ threshold. This result is important because it shows that the same QQP--DVQE solver interface can handle the LP case by setting $A=0$, without changing the main algorithmic structure.

For easy LPs, 24 out of 30 cases satisfy the $1\%$ quality rule. The objective and feasibility errors remain small: the maximum objective gap is only $0.084832\%$, and the maximum constraint gap is only $0.078025\%$. The lower success rate is mainly caused by the solution infinity-norm gap. This behavior is reasonable for LPs because linear programs can have nonunique or nearly nonunique optimal solutions. In such cases, two feasible solutions may have almost identical objective values but different decision-vector values. Therefore, the solution-vector gap should be interpreted together with the objective and constraint gaps, not in isolation.

Overall, the results validate QQP with DVQE as the internal QUBO solver. The experiment shows that the proposed framework can solve constrained continuous QP and LP instances by repeatedly forming QUBO subproblems and solving them with DVQE. The main result is solution quality: the objective gaps and constraint residuals remain far below $1\%$ for most cases, including the badly scaled hard QP and hard LP instances. The large runtime is expected in the current CPU-simulated implementation because each QQP iteration requires many DVQE-based QUBO solves. Thus, the timing results reflect the cost of repeated quantum-circuit simulation, while the quality results demonstrate that DVQE can serve as the QUBO backend inside QQP.

\subsubsection{Large-Scale Validation of QQP}
\label{subsubsec:qqp_large_scale_easy_qp}

The final experiment evaluates the large-scale behavior of QQP on higher-dimensional constrained QP instances. The purpose of this experiment is to test whether the QQP reformulation remains accurate when the number of continuous variables increases beyond the small instances used in the previous validation. Since the goal is to study the algorithmic scalability of QQP rather than the runtime of CPU-based quantum-circuit simulation, this experiment uses a classical local-search QUBO backend inside QQP. Thus, each local one-bit QUBO generated by the QQP refinement loop is solved by a classical random-restart bit-flip search. 

The tested QP instances have dimensions
\(
n\in\{50,75,100,150,200,300\}.
\)
Each instance is a strictly convex constrained QP with equality constraints, inequality constraints, and box bounds. The quality of the QQP solution is measured using the same percentage-based metrics as in the previous experiments. Table~\ref{tab:qqp_large_scale_easy_qp} reports the results.

\begin{table}[!t]
\centering
\caption{Large-scale QQP validation}
\label{tab:qqp_large_scale_easy_qp}
\scriptsize
\setlength{\tabcolsep}{3.5pt}
\renewcommand{\arraystretch}{1.10}
\begin{tabular}{cccccccc}
\toprule
$n$ & $m$ & $p$ & Obj. gap (\%) & $x$-gap (\%) & Constr. gap (\%) & AL iters & QUBO calls \\
\midrule
50  & 2  & 15 & 0.000298 & 0.003815 & 0.014500 & 3 & 1501 \\
75  & 4  & 22 & 0.000189 & 0.003732 & 0.019727 & 3 & 1245 \\
100 & 5  & 30 & 0.000184 & 0.002652 & 0.036041 & 3 & 1621 \\
150 & 8  & 45 & 0.000240 & 0.002926 & 0.013834 & 3 & 1787 \\
200 & 10 & 60 & 0.000183 & 0.003572 & 0.021689 & 3 & 1121 \\
300 & 15 & 90 & 0.000332 & 0.004513 & 0.040494 & 3 & 1234 \\
\bottomrule
\end{tabular}
\end{table}

The results show that QQP maintains high solution quality as the problem dimension increases. All six cases satisfy the $1\%$ quality rule. The objective gaps are extremely small, ranging from $0.000183\%$ to $0.000332\%$. The normalized solution infinity-norm gaps are also small, with the largest value equal to $0.004513\%$ for the $n=300$ case. The maximum normalized constraint gap remains below $0.05\%$ in all cases, which is far below the $1\%$ acceptance threshold.

The augmented-Lagrangian loop converges in only three iterations for every tested dimension. This indicates that the PHR fixed-region construction remains stable for these large-scale easy QP instances. The number of local QUBO calls varies between $1121$ and $1787$.

The reference solver reports successful termination for the $n=50$, $75$, $100$, and $200$ cases. For the $n=150$ and $n=300$ cases, the reference solver reports a positive directional derivative message, although the returned solutions have very small feasibility residuals. Therefore, these two reference solutions are used as numerical comparison points rather than certified optimal references. Overall, this experiment shows that the QQP reformulation can scale to QP instances with hundreds of continuous variables while preserving high solution quality.

\section{Conclusion}
\label{sec:conclusion}

This paper presented an open-source DVQE software framework for distributed QUBO solving and quadratic programming available at \href{https://github.com/LSU-RAISE-LAB/DVQE.git}{GitHub}. The proposed DVQE solver supports both monolithic and distributed quantum-circuit execution and evaluates QUBO objective values directly from measured bitstrings. To improve variational training, DVQE combines metaheuristic warm-start initialization with sampling-based variational refinement, where Black Hole optimization, Grey Wolf Optimization, and Artificial Bee Colony optimization are supported as warm-start strategies. This paper also developed QQP, a sequential QP to QUBO reformulation framework that uses variable scaling, augmented-Lagrangian fixed-region modeling, and repeated local one-bit QUBO reformulations to connect bounded constrained QP models with QUBO-based solvers.

Numerical experiments evaluated the proposed framework on QUBO and QP test problems. The results showed that distributed DVQE with Black Hole warm start achieved the strongest performance among the tested DVQE configurations, and that QQP can solve bounded constrained QP problems through the proposed QUBO-based workflow with scale-aware optimality, feasibility, and solution gaps. This work does not claim quantum speedup, since the experiments are performed using a CPU-based quantum-circuit software. Instead, the main contribution is a modular and testable open-source software framework that connects distributed variational quantum execution, QUBO optimization, and constrained quadratic programming.

\section*{Acknowledgments}

This work was supported by the U.S. National Science Foundation under Grant ECCS-1944752 and Grant ECCS-2312086.

\section*{Conflict of interest}

The authors declare that they have no conflicts of interest.

\section*{Ethical statement}

This work does not involve human participants, animal subjects, or clinical data.

\section*{Author contributions}

Milad Hasanzadeh developed the proposed DVQE and QQP software frameworks, implemented the numerical experiments, analyzed the results, and prepared the manuscript. Amin Kargarian supervised the research, guided the methodology, and reviewed and edited the manuscript.

\section*{Data and software availability}

The source code, tutorial examples, and experiment runner files associated with this work are publicly available at \url{https://github.com/LSU-RAISE-LAB/DVQE.git}. The numerical results reported in this paper can be reproduced using the scripts provided in the repository.

\bibliographystyle{ieeetr}
\bibliography{sn-bibliography}

\end{document}